\documentstyle[sprocl]{article}

\newcommand{\beq}{\begin{equation}}
\newcommand{\eeq}{\end{equation}}
\newcommand{\beqa}{\begin{eqnarray}}
\newcommand{\eeqa}{\end{eqnarray}}
\newcommand{\Adir}{\mbox{${\cal A}^{\rm dir}_{\rm CP}$}}
\newcommand{\bBz}{\mbox{${\overline{B^0}}$}}
\newcommand{\bKz}{\mbox{${\overline{K^0}}$}}

\def\gev{\,{\rm GeV}}

\begin{document}

\begin{flushright}
JHU--TIPAC--98008\\
hep-ph/9806538\\
July, 1998
\end{flushright}

\vspace{1cm}

\title{FINAL STATE INTERACTIONS IN HADRONIC $B$ DECAYS\footnote{To appear in the proceedings of {\it Continuous Advances in QCD 1998,} Theoretical Physics Institute, University of Minnesota, April 16-19, 1998.}}

\author{ADAM F.~FALK}

\address{Department of Physics and Astronomy\\
The Johns Hopkins University\\
3400 North Charles Street\\
Baltimore, Maryland 21218 U.S.A.\\
{\tt falk@jhu.edu}}

\maketitle\abstracts{I discuss the effect of final state interactions on the determination of the CKM angle $\gamma$ from $B\to K\pi$ decays.  Using a simple Regge-based model for rescattering processes, I argue that such effects could be substantial enough to make it problematic to obtain reliable limits on $\gamma$ in this way.  Fortunately, an analysis of $B\to KK$ decays may provide model-independent bounds on rescattering contributions.}

\section{Features of $B\to K\pi$ decays}

The decays $B\to\pi^\mp K^\pm$ and $B^\pm\to\pi^\pm K$ have attracted considerable recent interest,\cite{FMbound} because of the insight their study might provide into bounding the CKM parameter $\gamma$ or finding a clean signal of new $CP$ violating physics.  However, the success of this program depends on the assumption that one can use na\"\i ve arguments based on quark flow diagrams to determine the weak phases on which these decays depend.  Such arguments typically neglect inelastic final state rescattering processes.  Recently, we have studied the effect of final state interactions on the weak phase dependence of the matrix elements relevant to $B\to K\pi$, and, in particular, on the proposed bounds on $\sin\gamma$.\cite{FKNP}  Other authors have made similar investigations, using a variety of approaches.~\cite{GR,BpiK,rescatt}

In the Standard Model, the decays $B\to\pi^\mp K^\pm$ and $B^\pm\to\pi^\pm K$ are mediated by operators of the ``current-current'' type and of the ``penguin'' type.  While this distinction is a popular one, it is neither renormalization group invariant nor rigorously defined.  For our purposes, it is most useful to decompose the amplitudes which govern these processes according to their dependence on $\gamma$,
\beqa \label{Camps}
  &&A(B^+ \to \pi^+ K^0) = A_{cs}^+  - A_{us}^+ e^{i \gamma} e^{i \delta_+},
  \nonumber\\
  &&A(B^- \to \pi^- \bKz) = A_{cs}^+ - A_{us}^+ e^{-i \gamma} e^{i \delta_+},
  \nonumber\\
  &&A(B^0 \to \pi^- K^+) =  A_{cs}^0 - A_{us}^0 e^{i \gamma} e^{i \delta_0},
  \nonumber\\
  &&A(\bBz \to \pi^+ K^-) =  A_{cs}^0 - A_{us}^0 e^{-i \gamma} e^{i \delta_0},
\eeqa
where $\delta_0$ and $\delta_+ $ are $CP$ conserving phases induced by the strong interaction.  One can then use the decomposition into ``penguin'' and ``tree'' amplitudes to make a {\it na\"\i ve} estimate of the sizes of the $A$'s.  One finds that in $B^0$ decays, $A^0_{cs}$ is a penguin decay and is of order $\lambda^2/16\pi^2$, where $\lambda=\sin\theta_C\approx0.2$, while $A^0_{us}$ is a tree decay of order $\lambda^4$.  Hence the two amplitudes which contribute to $B^0 \to \pi^- K^+$ are roughly of the same size.  By contrast, there is no tree contribution to $B^+ \to \pi^+ K^0$, and both $A^+_{cs}\sim\lambda^2/16\pi^2$ and $A^+_{us}\sim\lambda^4/16\pi^2$ come from penguin graphs.  (This is because $B^+ \to \pi^+ K^0$ arises from a valence quark transition of the form $b\to d\bar ds$, which, unlike $b\to u\bar us$, is not mediated by a tree level weak decay.)  Hence one na\"\i vely expects $A^+_{us}$ to be negligible compared to $A^+_{cs}$, a result which, if true, would have interesting consequences.

The first of these comes from the $CP$ violating asymmetry
\beq\label{ABpiK}
  \Adir={
  B(B^+\to\pi^+K^0)-B(B^-\to\pi^-\bKz)\over
  B(B^+\to\pi^+K^0)+B(B^-\to\pi^-\bKz)}\,.
\eeq
In the Standard Model, if these estimates are correct then $\Adir=O(\lambda^2)$, and the observation of a substantial $\Adir$ would be a clean signal of new $CP$ violating physics.  Second, one can define the $CP$ conserving ratio
\begin{equation} \label{ratio}
  R = \frac{B(B^0 \to \pi^- K^+) + B(\bBz \to \pi^+ K^-)}{
  B(B^+ \to \pi^+ K^0) + B(B^- \to \pi^- \overline{K^0})}\,,
\end{equation}
and use it to derive the bound $\sin^2\gamma\le R$.\cite{FMbound}  Since current data yield $R=0.65\pm0.040$, this limit could prove interesting as the experimental situation improves.\cite{CLEO}

Crucial to both of these results is the estimate $A^+_{us}\sim\lambda^4/16\pi^2$.  This estimate relies on the assumption that $A^+_{us}$ is dominated by a penguin operator, generated by a $u-W$ loop at short distances.  However, long-distance contributions are also possible, in which an intermediate $u$ and $\bar u$ go on shell.  Computed perturbatively, one finds only a small effect.\cite{BSS}  However, for small loop momenta the duality approximation fails, in that it is inappropriate to have on-shell light quarks, and one must consider on-shell mesons instead.  In such a case, this is really a rescattering process such as $B^+\to K^+(n\pi)\to K\pi^+$.  One can argue about whether such a process has the ``topology'' of a tree or a penguin diagram, but the distinction is not a meaningful one.  What is important is that rescattering could actually give the dominant contribution to $A^+_{us}$. The point is that for a process like $B^+\to K^+\pi^0\to K^0\pi^+$, the ``penguin'' amplitude is not suppressed by a perturbative short distance loop factor $1/16\pi^2$; instead, its magnitude is controlled by nonperturbative hadronic physics.

We will be interested in the possibility that such final state interactions (FSI) make $A_{us}^+$ much larger than our na\"\i ve estimate.  Let us define $\epsilon=A^+_{us}/A^+_{cs}$, and keep terms linear in $\epsilon$.  Then we find the modified asymmetry
\begin{equation}
  \Adir=2\epsilon\sin\gamma\sin\delta_+\,,
\end{equation}
and the modified bound
\beq\label{fmboundnew}
  \sin^2\gamma\le R(1+2\epsilon\sqrt{1-R})\,.
\eeq
Suppose $|\epsilon|\simeq10\%$.  Then, in contrast to our earlier estimate, the observation $|\Adir|\sim20\%$ would {\it not} be an unambiguous signal of new physics.  Furthermore, the bound on $\gamma$ would be severely degraded.

\section{A model for two-body rescattering}

\subsection{Warning and disclaimer}

To explore the magnitude of rescattering effects which one might expect, we propose a simple model of two body rescattering based on Regge phenome\-nol\-ogy.\cite{Regge}  A few points are in order.  First, we have little confidence in the quantitative predictions of the model {\it per se.}  In fact, we believe that it would be irresponsible to claim an accuracy of better than a factor of two for the size of soft rescattering effects, using {\it any\/} model currently available.  Neither this nor any other model should be taken as a canonical framework for the estimate of final state interactions in $B$ decays.  Rather, the purpose of our calculation is to be illustrative: our model will predict $\epsilon$ at the level of ten percent, with no fine tuning or unnatural enhancements.

However, our model does have certain virtues with respect to other recent estimates of final state interactions.\cite{GR,BpiK,rescatt}  First, we go beyond a simple model of elastic final state interaction phases.  While an elastic treatment is adequate for $K\to\pi\pi$ transitions, inelastic rescattering is dominant for $B$ decays.  (In fact, even our model is unable to go far enough in this respect, since we will include only two-body intermediate states, and it is really multibody intermediate states which are the most important.\cite{Regge})  Second, by eschewing a description in terms of ``quark flow'' diagrams, we avoid relying implicitly on the very same na\"\i ve estimates whose validity it is our purpose to explore.  Once again, the issue of whether the processes we consider have a ``tree'' or a ``penguin'' topology is a dangerous red herring: the question has no scale-invariant meaning, and also no important implications.  Our model simply addresses contributions to the well-defined amplitude $A^+_{us}$.

To reiterate, we {\it do not\/} expect an accurate computation of final state interactions from this model.  Nor do we trust any other model to provide one.  What we {\it do\/} expect is to estimate the {\it generic size\/} of rescattering effects.  Hence, we will argue that the common assumption that final state interactions are substantially {\it smaller\/} than what we find is not a generic one, and would require some concrete and substantial justification.

\subsection{The model}

The model which we will use is purely phenomenological, and is based on the exchange of Regge trajectories.  It is a feature of hadronic physics that total hadronic cross sections are fit well by an expression of the form
\begin{equation}\label{regge}
  \sigma_{ij}(s)=X_{ij}\left({s/ s_0}\right)^{0.08}+
  Y_{ij}\left({s/ s_0}\right)^{-0.56}\,,
\end{equation}
where $s_0=1\gev$ is a scale typical of the strong interactions, and $i$ and $j$ refer to the particular initial states.  The exponents are observed to be universal.  The two terms in Eq.~(\ref{regge}) may be interpreted as arising from the exchange of distinct Regge trajectories.  The first comes from Pomeron exchange, which is responsible for the asymptotic dependence of the total hadronic cross section.  The second comes from the exchange of subleading trajectories such as the $\rho$ and the $a_1$.  Note that what is exchanged in a trajectory is not a finite number of individual resonances, which could not lead to the observed $s$-dependence, but rather an entire family of states.  

Because of its soft dependence on $s$, the cross section (\ref{regge}) is not negligible for $s\sim m_B^2$.  One might be tempted to conclude from this that {\it re\/}scattering of the form $B\to K\pi\to K\pi$ is substantial as well.  However, one must be careful, because the Regge parameterization for the total cross section is applicable to peripheral partial waves, at large values of $\ell$.  By contrast, $B\to K\pi$ proceeds through the $\ell=0$ channel, and can receive contributions from Regge cuts.\footnote{I am grateful to J.~Rosner for discussions of this point.}  This drawback of the model affects our ability to derive precise predictions from it, as does the fact that we will be forced to neglect multibody intermediate states such as $K(n\pi)$.  At this point, we recall that our goal is simply to estimate the generic magnitude of the rescattering contribution, and proceed with these caveats in mind.

We are interested in the rescattering processes $B^+\to K^+\pi^0\to K^0\pi^+$ and $B^+\to K^+\eta\to K^0\pi^+$, in which quark flavor quantum numbers are exchanged. The leading Regge trajectory is the exchange of the Pomeron, which carries no flavor; hence it cannot contribute here. Instead, the leading contribution will come from the subleading trajectory, which includes the charged $\rho$.  We will use $SU(3)$ flavor symmetry in our estimate, so this trajectory contains the $K^*$ as well.  The error made by applying $SU(3)$ is usually on the order of $30\%$, which is small compared to the uncertainties discussed above.

The details of the rest of the calculation are found elsewhere;\cite{FKNP,Regge} here I will simply outline the procedure, highlighting the most important physics points.  Since there are no actual data on $K\pi$ scattering, the residue $Y_{K\pi}$ must be extracted from fits to $p\pi$ and $pp$ data, plus flavor $SU(3)$.  Then we find $Y_{K\pi}\propto Y_{\pi p}^2/Y_{pp}$, with different Clebsch-Gordan coefficients for the four relevant channels: $K^+\pi^0\to K^0\pi^+$ and $K^+\eta\to K^0\pi^+$, each via exchange of charged $\rho$ and $K^*$ trajectories.

Let us focus for concreteness on $K^+\pi^0\to K^0\pi^+$ via $\rho^+$ exchange.  Since this is a contribution to $\epsilon$ from the two-body intermediate state, we will denote it by $\epsilon_2$.   We begin by writing an expression for the discontinuity of the amplitude for the charged $B$ decay,
\begin{eqnarray} \label{disc}
  {\rm Disc} \,A (B^+ \to {K^0} \pi^+)_{\rm FSI} =&&\!\!\!\!\! \frac{1}{2}
  \int [{\rm d}p_1] [{\rm d}p_2]\,
  (2 \pi)^4 \delta (p_B - p_1 - p_2)
  \\
  &&\!\!\!\!A (B^+ \to \pi^0 (p_1) K^+(p_2))
  \cdot{\cal M} (\pi^0 K^+\to {K^0} \pi^+)\,.\nonumber 
\end{eqnarray}
The rescattering matrix element ${\cal M} (\pi^0 K^+\to{K^0} \pi^+)$ has a well known parameterization inspired by Regge
phenomenology, which for the exchange of the $\rho$ trajectory
may be written as
\begin{eqnarray} \label{reggeme}
  {\cal M} (\pi^0K^+ \to {K^0} \pi^+) =
  \gamma(t)\ \frac{e^{-i \pi\alpha (t)/2}}{
  \cos (\pi\alpha (t)/2)}
  \ \left ( \frac{s}{s_0} \right )^{\alpha (t)}\,,
\end{eqnarray}
where $\gamma(t)$ is a residue function and
$s$ and $t$ are the Mandelstam variables.  We take a linear Regge trajectory,
$\alpha(t)=\alpha_0+\alpha't$, with $\alpha_0=0.44$ and
$\alpha'=0.94\,{\rm GeV}^{-2}$ for $\rho$ exchange.
Taking $\gamma(t)=\gamma(0)\equiv\gamma$ for simplicity, the discontinuity
can be calculated,
\begin{equation}
  {\rm Disc} \,A (B^+ \to {K^0} \pi^+)_{\rm FSI} =  
  \gamma\,\bar\epsilon_2(s)\,
  \left(s/s_0\right )^{\alpha_0 - 1}
  A (B^+ \to \pi^0  K^+)\,,
\end{equation}
where
\begin{equation}
  \bar \epsilon_2 (s) = \frac{1}{16 \pi}\
  \frac{1}{\cos (\pi\alpha_0/2)}\
  \frac{e^{-i \pi\alpha_0/2}}{s_0\alpha'
  \left( \ln (s/s_0) - i \pi /2 \right)}\,.
\end{equation}
We restore the FSI contribution to
$A (B^+\to {K^0} \pi^+)$ by use of a dispersion relation.  The dispersion
integral may be evaluated in closed form with the approximations
$\alpha_0={1\over2}$ and $\ln(s/s_0)=\ln(m_B^2/s_0)$,
\begin{eqnarray}
  A (B^+ \to {K^0} \pi^+)_{\rm FSI} &=&
  \gamma\,\bar \epsilon_2 (m_B^2)\,A (B^+ \to
  \pi^0 K^+)\cdot {1\over\pi} \int
  \frac{{\rm d}s}{s-m_B^2} \left ( \frac{s}{s_0} \right )^{\alpha_0 - 1}
  \nonumber \\
  &=& i\,\gamma\,\bar \epsilon_2 (m_B^2)\, \frac{\sqrt{s_0}}{m_B}\,
  A (B^+ \to \pi^0 K^+)\,,
\end{eqnarray}
where the integral runs over $0\le s\le\infty$ in the limit $m_\pi=m_K=0$.  With the same approximations, we find the residue $\gamma=\gamma_{\pi K\rho}=-\sqrt{2}\,s_0 Y_{\pi p}^2/Y_{pp}\approx 51$, using data from the Particle Data Group.\cite{PDG}
For rescattering through other channels, the only difference is a
Clebsch-Gordan coefficient in the residue function $\gamma$.  The magnitude
of the contribution of a given channel to the soft rescattering amplitudes
is then
\begin{equation}\label{eps2}
  \epsilon_2={\sqrt{s_0}\over m_B}\,|\gamma\,\bar\epsilon_2(m_B^2)|\,,
\end{equation}
where $\gamma$ depends on the channel: $\gamma_{\eta K\rho}=0$, $\gamma_{\pi KK^*}=(1/2)\gamma_{\pi K\rho}$ and $\gamma_{\eta KK^*}=-(\sqrt{3}/2)\gamma_{\pi K\rho}$.

Finally, we need an estimate of the amplitude to produce the intermediate state, $A (B^+ \to \pi^0 K^+)$, or rather we require the ratio $A (B^+ \to \pi^0 K^+)/A_{cs}^+$.  For this ratio, we rely on the BSW model,\cite{BSW} supplemented by $SU(3)$ symmetry.  Admittedly, this is a crude treatment of these matrix elements, but given the large uncertainties elsewhere in our calculation, and the fact that we use the BSW model to compute a ratio rather than an absolute rate, it should be sufficient for our purposes.  Leaving the details to elsewhere,\cite{FKNP} we find $|A (B^+ \to \pi^0 K^+)/A^+_{cs}|\simeq0.35$ and $|A (B^+ \to \eta K^+)/A^+_{cs}|\simeq0.20$.

\subsection{The result}

Combining the contributions from the different channels, we arrive at the estimate~\cite{FKNP}
\begin{equation}
  \epsilon\sim0.05 - 0.10\,.
\end{equation}
It is useful to pause at this point to recall some of the sources of uncertainty in this determination.  In roughly decreasing order of importance, they include:
\par 1.\ The inclusion only of two-body intermediate states.
\par 2.\ The use of Regge phenomenology for $\ell=0$ partial wave.
\par 3.\ The unknown relative strong phases of the different channels.
\par 4.\ The neglect of the intermediate state $\eta'K^+$.
\par 5.\ The use of the BSW model.
\par 6.\ The use of $SU(3)$ flavor symmetry.
\par\noindent In light of all these uncertainties, it would be wildly optimistic to hope for an accuracy in $\epsilon$ of better than a factor of two.  With that in mind, $\epsilon\sim0.1$ would be a reasonable value, and even $\epsilon\sim0.2$ would not be unreasonable.  We note that there are certainly models in which $\epsilon$ is found to be smaller than this.  However, unless such a model is more reliable than ours, {\it in the sense that it includes physics effects which this one does not,} it cannot by itself give evidence that the value of $\epsilon$ found here is generically too large.  In particular, this comment applies to all models which parameterize FSI only with arbitrary elastic final state phases.

Values of $\epsilon$ such as we have found have potentially profound phenomenological effects.  First, an observation of $|\Adir|$ as large as 20\% or 30\% would not be an unambiguous signal of $CP$ violating physics beyond the Standard Model.  Second, the bound (\ref{fmboundnew}) would be severely degraded, yielding little or no information on $\sin\gamma$ unless $R$ were measured to be significantly less than 1 and with very small errors.

\section{Experimental bounds on final state interactions}

In view of the possibility of large rescattering contributions to $B^+\to K^0\pi^+$, it would be extremely useful to be able to bound the FSI contribution from some experimental information.  Fortunately, this is possible.  The idea is to study decay modes mediated by the quark level transition $b\to s\bar sd$, which is related by $SU(3)$ symmetry to the $b\to d\bar ds$ transition of interest to us.  (More precisely, it is related by $U$-spin, the $SU(2)$ subgroup of $SU(3)$ which exchanges $s\leftrightarrow d$.)  Then branching ratio measurements or upper bounds on the new modes would imply a direct upper bound on $\epsilon$.

The most useful modes turn out to be $B^+\to \overline{K^0} K^+$.  With the inclusion of rescattering processes which at the quark level are of the form $b\to u\bar ud\to s\bar sd$, the amplitudes may be decomposed in the form
\begin{eqnarray}
  &&A(B^+\to\overline{K^0} K^+) = A_{cd}-A_{ud}e^{i \gamma}e^{i\delta},
  \nonumber\\
  &&A(B^-\to{K^0} K^-) = A_{cd}-A_{ud}e^{-i \gamma}e^{i\delta}.
\end{eqnarray}
Recall that in $B^+\to K^0\pi^+$, we suppose that the penguin amplitude $A_{cs}^+$ is dominant, while the rescattering contribution $A_{us}^+$ is smaller by the factor $\epsilon$.  We now use $SU(3)$ to relate $A_{cs}^+$ and $A_{us}^+$ to $A_{cd}$ and $A_{ud}$.  We find
\begin{eqnarray}
  &&\left|{A_{cd}/A_{cs^+}}\right|=\left|V_{cd}/V_{cs}\right|=\lambda\,,
  \nonumber\\
  &&\left|{A_{ud}/A_{us^+}}\right|=\left|V_{ud}/V_{us}\right|=1/\lambda\,;
\end{eqnarray}
that is, the relative size of the rescattering amplitude in $B^+\to \overline{K^0} K^+$ is not $\epsilon$ but rather $\epsilon/\lambda^2\sim20\epsilon$.  Hence rescattering contributions could easily dominate the decay rate in this channel, and assuming no fine tuned cancelations, one can derive bounds on $\epsilon$ by studying this process.

The experimental bound is derived by measuring the $CP$ conserving ratio
\begin{equation}
  R_K = \frac{B(B^+ \to \overline{K^0} K^+) + B(B^- \to K^0 K^-)}
  {B(B^+ \to \pi^+ K^0) + B(B^- \to \pi^- \overline{K^0})}\,.
\end{equation}
Then, to leading order in $\lambda$, we find
\begin{equation}
  \epsilon<\lambda\sqrt{R_K}+O(\lambda^2)\,.
\end{equation}
Inserting this into the bound on $\sin\gamma$ yields
\begin{equation}\label{gambound}
  \sin^2\gamma<R\left(1+2\lambda\sqrt{R_K(1-R)}\right)+O(\lambda^2)\,.
\end{equation}
For the $CP$ violating asymmetry $\Adir$, the result depends on whether the new bound (\ref{gambound}) is nontrivial.  Without information from (\ref{gambound}), we find
\begin{equation}
  |\Adir|<2\lambda\sqrt{R_K}+O(\lambda^3)\,,
\end{equation}
while if we can incorporate the bound on $\sin^2\gamma$, we obtain
\begin{equation}
  |\Adir|<2\lambda\sqrt{R_K R}+O(\lambda^2)\,.
\end{equation}

The CLEO Collaboration has obtained the upper bound $R_K < 0.95 $ at 90\% c.l., including only statistical errors, and approximately $R_K < 1.9 $ once systematic errors are included as well.\cite{CLEO,GW}   This is not yet very restrictive, implying $\epsilon < 0.4 $ and $\Adir < 0.6 $.  More interesting constraints on $\epsilon$ and $\Adir$ possibly could be obtained in the future, given that $R_K \sim {\cal O}(\lambda^2)$ in the limit of vanishing FSI. Ultimately, the utility of such a bound will be limited by the fact that, for $\epsilon$ small enough, $R_K$ becomes independent of $\epsilon$, since rescattering channels then would be negligible compared with other contributions.

\section{Summary}

This has been an investigation into the effect of final state interactions on the decay mode $B\to K\pi$.  While it is certainly true that they have a small effect on the total $B$ decay rate, as one might expect, this is not necessarily true for individual decay channels.  In particular, rescattering processes, even if relatively small, can have a phenomenologically important effect on the CKM structure of transition amplitudes.  In the case at hand, rescattering effects of order 10\% to 20\% can spoil bounds on $\sin\gamma$ and signals for new physics which otherwise could be derived from $B\to K\pi$ decays.

We have no accurate model for calculating the size of final state interactions.  Instead, we have employed a phenomenological Regge-based approach, which we believe gives information about the generic size of rescattering effects.  We trust this model only in the sense that it serves to highlight the large uncertainties associated with these processes.  While it is quite possible that rescattering contributions are smaller than we have estimated here, one cannot simply rely on this assumption if one hopes to explore $CP$ violation in the $B$ system reliably and realistically.  In particular, the existence of simpler models in which the rescattering is smaller and under better control, such as those based only on elastic strong interaction phases, is irrelevant to establishing confidence in the viability of this approach for probing fundamental physics.

\section*{Acknowledgements}

It is a pleasure to thank the organizers for once again arranging a very stimulating and enjoyable workshop.  This work was supported by  the National Science Foundation under Grant No.~PHY-9404057 and National Young Investigator Award No.~PHY-9457916; by the Department of Energy under Outstanding
Junior Investigator Award No.~DE-FG02-94ER40869; and by the Alfred
P.~Sloan Foundation.  A.F.~is a Cottrell Scholar of the Research Corporation.

\end{document}